\documentclass{article}
\usepackage{spconf,amsmath,graphicx,amsfonts,amssymb,hyperref}
\usepackage[symbol]{footmisc}

\makeatletter
\newcommand{\printfnsymbol}[1]{%
  \textsuperscript{\@fnsymbol{#1}}%
}
\makeatother

\title{DGC-vector: A new speaker embedding for zero-shot voice conversion}
\name{Ruitong Xiao\textsuperscript{1}\sthanks{  $ $ Equal contribution}\sthanks{ $ $  Work done during the internship at Netease Games AI Lab }, Haitong Zhang\textsuperscript{2}\printfnsymbol{1},  Yue Lin\textsuperscript{2} }

\address{\textsuperscript{1}South China University of Technology, China;\textsuperscript{2} Netease Games AI Lab, China  \\
\textsuperscript{1}eerxiao@mail.scut.edu.cn; \textsuperscript{2}\{zhanghaitong01, gzlinyue\}@corp.netease.com}
\begin{document}
\ninept
\maketitle
\begin{abstract}
Recently, more and more zero-shot voice conversion algorithms have been proposed. As a fundamental part of zero-shot voice conversion, speaker embeddings are the key to improving the converted speech's speaker similarity. In this paper, we study the impact of speaker embeddings on zero-shot voice conversion performance. To better represent the characteristics of the target speaker and improve the speaker similarity in zero-shot voice conversion, we propose a novel speaker representation method in this paper. Our method combines the advantages of D-vector, global style token (GST) based speaker representation and auxiliary supervision. Objective and subjective evaluations show that the proposed method achieves a decent performance on zero-shot voice conversion and significantly improves speaker similarity over D-vector and GST-based speaker embedding.
\end{abstract}


\section{Introduction}
Voice conversion (VC) task is to convert a source speaker's voice to a target speaker's voice without changing the linguistic content~\cite{BerrakTASLP2021}. It is widely applied in many fields such as entertainment, creative industry, and spoofed speech generation. 

In terms of learning methods, previous works of VC can be roughly divided into two categories: automatic speech recognition (ASR) based approach and auto-encoder based approach. The ASR-based VC first extracts the phoneme posteriorgram (PPG) from the utterance by a pre-trained ASR model as the content features and then performs voice conversion using PPG. PPG is considered a kind of speaker-agnostic feature. ASR-based VC methods, such as \cite{sun2016phonetic}, are widely used, but the performances highly depend on the performances of ASR models. Different from ASR-based VC, auto-encoder-based VC methods, such as \cite{hsu2016voice}, extract the content features through the content encoder and then perform voice conversion on the content features. Auto-encoder-based VC systems need to disentangle the speaker identity from the content information. Besides that,  StarGAN-VC~\cite{Kaneko2019StarGAN} and CycleGAN-VC~\cite{Kaneko2019CycleGAN} are two typical methods.
Most of the previous VC works can perform VC on the seen speaker data, namely seen-to-seen VC, but are infeasible to few-shot VC. Thus, their scope of application is restricted. Few-shot VC is more challenging than seen-to-seen VC and has become a hot spot of current research. To overcome the restraint, some methods with a strong generalization ability have been proposed for the few-shot or even zero-shot VC task.

In the case of few-shot VC, speaker embedding is an important factor that can not be ignored since it can help VC models use limited training data more efficiently and generate voice more similar to the target speaker. Usually, the speaker embedding is generated by a pre-trained speaker encoder or a jointly-optimized speaker encoder.
In Voice Conversion Challenge (VCC) 2020~\cite{zhao2020voice}, two kinds of speaker embedding are widely used. Some teams like \cite{zhang2020neteasegames} and \cite{mayor2020fastvc} simply used the one-hot embedding as the speaker embedding.  Automatic speaker verification (ASV) based speaker embedding was used in many schemes~\cite{ma2020submission, liu2020non}.

Compared with few-shot VC, zero-shot VC algorithms do not need training the VC model with data from the target speaker. Therefore, the speaker embedding of zero-shot VC algorithms should have the ability to represent the characteristics of the target speaker. For example, VQVC~\cite{9053854} proposes to separate the speaker and content information through discrete codes and uses the residual feature for speaker representation. AUTOVC~\cite{Qian2019ZeroShotVS} uses D-vector~\cite{variani2014deep,wan2018generalized} as the speaker embedding. Chou et al. propose a zero-shot voice conversion system with instance normalization (AdaIN-VC)~\cite{Chou2019One}, which jointly optimizes a speaker encoder with the conversion model.

In this paper, we study the impact of speaker embeddings on zero-shot VC performance and aim to improve speaker similarity of zero-shot VC. In order to represent the characteristics of target speakers better and improve the speaker similarity, we propose a novel speaker representation method, namely DGC-vector. Our method combines the advantages of D-vector, global style token~\cite{wang2018style} based speaker representation and auxiliary supervision. A GST-based module with auxiliary speaker classifier is applied to generate speaker embedding from bottleneck feature of D-vector extractor. Objective and subjective evaluations show that our proposed method achieves a decent performance on zero-shot voice conversion and improves the speaker similarity significantly over other baseline speaker representations.

The rest of the paper is organized as follows. Section 2 simply introduces related works. Section 3 introduces the proposed method in detail. Section 4 gives the experimental results and the corresponding analysis. Finally, the paper is concluded.

\section{Related work}

\subsection{D-vector} \label{d-vec}

% D-vector~\cite{variani2014deep} is a typical ASV based speaker embedding as shown in part (a) of Fig.~\ref{SE}. Usually, the extractor is trained for speaker classification task or clustering task. Similar ASV based speaker embedding include x-vector, i-vector and so on. Furthermore, Generalized end-to-end (GE2E)~\cite{wan2018generalized} method was proposed. It is a contrastive learning method and is widely applied in both VC and multi-speaker text-to-speech (TTS) systems as a well-trained separate part. For example, in Fig.~\ref{autovc}, AUTOVC takes D-vector (trained with GE2E loss) as speaker embedding $z_{s}$ to represent speaker characteristics. 

%%%D-vector extractor, which is consist of long short-term memory (LSTM) and fully connected (FC) layers, takes mel-spectrograms as input and produces D-vector from last frame of LSTM layers output by a FC layer. 

D-vector~\cite{variani2014deep} is a typical kind of speaker embedding extracted from the deep neural network based ASV system. Due to its outstanding performance in speaker verification, D-vector has been widely used to build a speaker adaptive model for both VC and text-to-speech (TTS) \cite{Qian2019ZeroShotVS, jia2018transfer}. \cite{wan2018generalized} proposes a generalized end-to-end loss for speaker verification, which achieved the SOTA results in speaker verification. The proposed system has been applied in some VC systems, such as AUTOVC \cite{Qian2019ZeroShotVS}. 

Regarding speaker verification, D-vector has proved to be generalizable to the unseen speakers to some extent \cite{wan2018generalized}. However, speaker embedding extracted from speaker verification systems may not adapt to few-shot or zero-shot VC well, leading to a poor speaker similarity between the target voice and the converted ones. Fig.~\ref{dv-tsne} shows the 2D t-SNE visualization of speaker embeddings computed from target speakers' ground truth utterances and converted utterances. The first left column corresponds to AUTOVC conditioned on D-vector extracted from \cite{wan2018generalized}. We define converted utterances and their corresponding target speakers' utterances as a conversion group. As shown in the upper part of the first left column in Fig.~\ref{dv-tsne}, the converted utterances are close to their target speakers' utterances, and different conversion groups are well separated. It indicates that D-vector adapts to AUTOVC well in seen-to-seen VC. However, as shown in the bottom part, some converted utterances are far from target speakers' utterances in zero-shot VC. It illustrates that the speaker similarity significantly decreases in the zero-shot conversion, and the model does not adapt well to the unseen speakers.

\subsection{Global Style Token}
Global style token (GST)~\cite{wang2018style} is able to extract a sentence-level representation from the reference utterance. It consists of a reference encoder and a style token module. In the original paper~\cite{wang2018style}, GST is used to capture the speaking style. In the M2VoC challenge~\cite{xie2021multi}, Wang et al.~\cite{wang2021prosody} used GST to represent speaker and prosody characteristics. For VC, Lu et al.~\cite{lu2019one} proposes to use the GST module as a speaker encoder. In this paper, we refer this kind of speaker embedding in ~\cite{lu2019one} as G-vector.
G-vector can adapt well the VC model since the GST module is jointly
optimized with the whole VC model. However, G-vector is extracted without any explicit supervision. Moreover, the GST module is trained with only the VC training dataset instead of a large multi-speaker dataset typically used in training ASV systems. Thus, the generalization ability of the G-vector may be limited.

\section{Proposed Method}

% In this section, we introduce a GST based speaker embedding method. We use AUTOVC as framework. Because we just focus on the impact of speaker embedding, so we do not change the setting of content encoder and decoder. Under the framework of AUTOVC, our method combine D-vector and GST to get speaker representation.

In this paper, we use AUTOVC~\cite{Qian2019ZeroShotVS} as the framework. Because we focus on the impact of speaker embedding, we keep the content encoder and decoder settings unchanged. Under the framework of AUTOVC, we propose a novel speaker embedding extraction method, which combines the advantages of the ASV-based method and the GST-based method. 

\subsection{Framework}
\begin{figure}[t]
\centering
\includegraphics[width=0.5\textwidth]{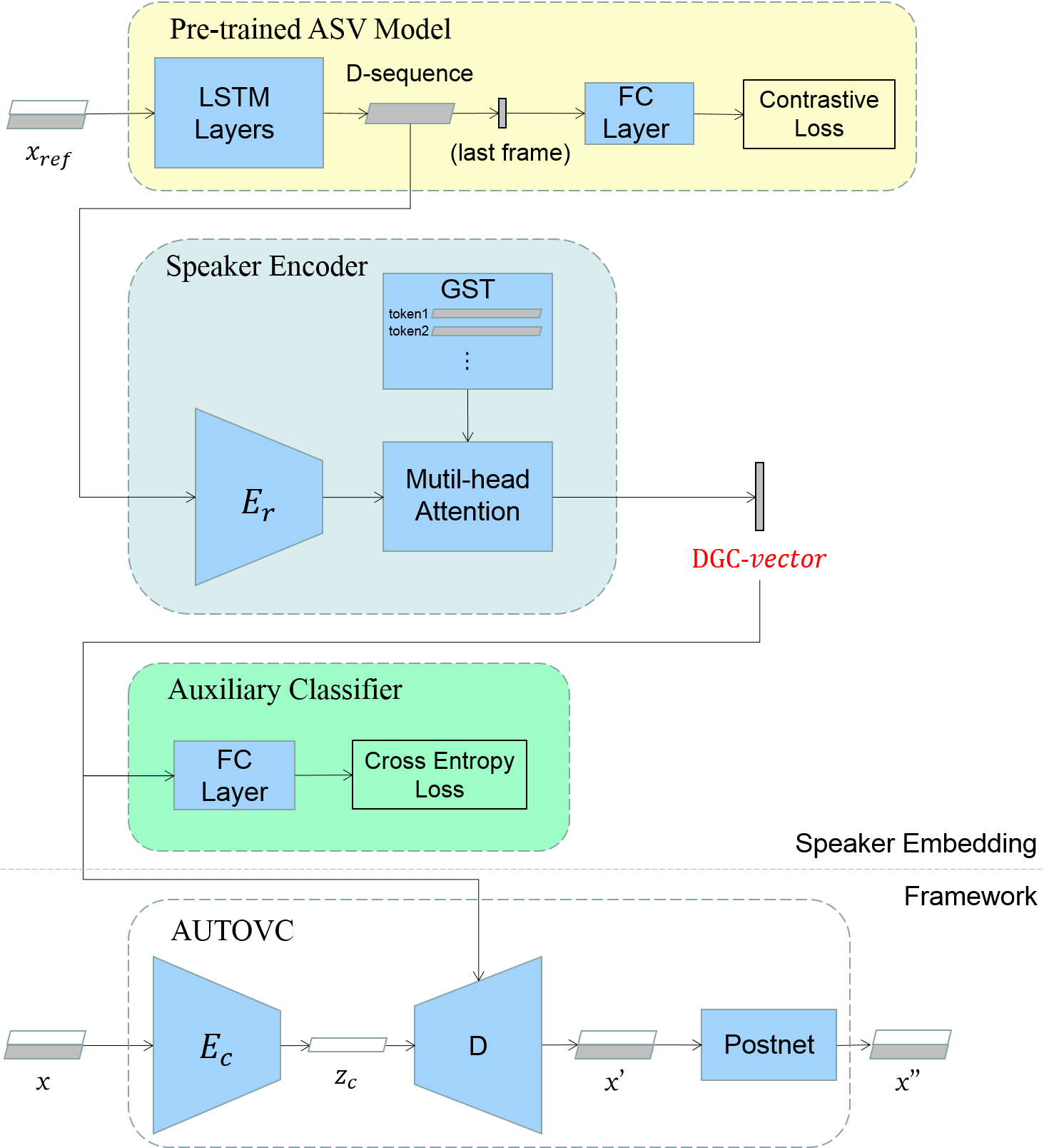} 
\caption{The architecture of entire model. Proposed speaker embedding method is shown in the upper part. The fixed LSTM layers extracts D-sequence as input of speaker encoder and then speaker encoder extracts speaker embedding, namely DGC-vector.}
\label{SE}
\end{figure}

\iffalse
\begin{figure}[t]
\centering
\includegraphics[width=0.45\textwidth]{fig/autovc.png} 
\caption{The training phase of AUTOVC. $x$, $z_{c}$ and $z_{s}$ represent input mel-spectrogram, content represenrarion and speaker embedding, respectively.}
\label{autovc}
\end{figure}
\fi
AUTOVC~\cite{Qian2019ZeroShotVS} is an auto-encoder VC model with a carefully designed bottleneck (See framework part of Fig.~\ref{SE}). AUTOVC is trained with the self-reconstruction loss like CVAE, but it has the distribution matching property similar to GAN. The loss function of AUTOVC is defined as

\begin{equation}
L_{rec}=\left\Vert x-x'\right\Vert _{2}+\left\Vert x-x''\right\Vert _{2}+\left\Vert E_{c}(x)-E_{c}(x'')\right\Vert _{1}
\end{equation}
where $E_{c}$ denotes content encoder. 
The fundamental idea of AUTOVC is that the correctly-designed bottleneck learns to disentangle the content information and speaker information by introducing carefully tuned dimension reduction and temporal downsampling. This simple scheme leads to a significant performance improvement. As a result, AUTOVC achieves superior performances on the many-to-many VC task and decent performances on the zero-shot VC task.

\begin{figure*}
\centering
\begin{minipage}{0.2\textwidth} \label{ssd}
\centering
\includegraphics[width=1\textwidth]{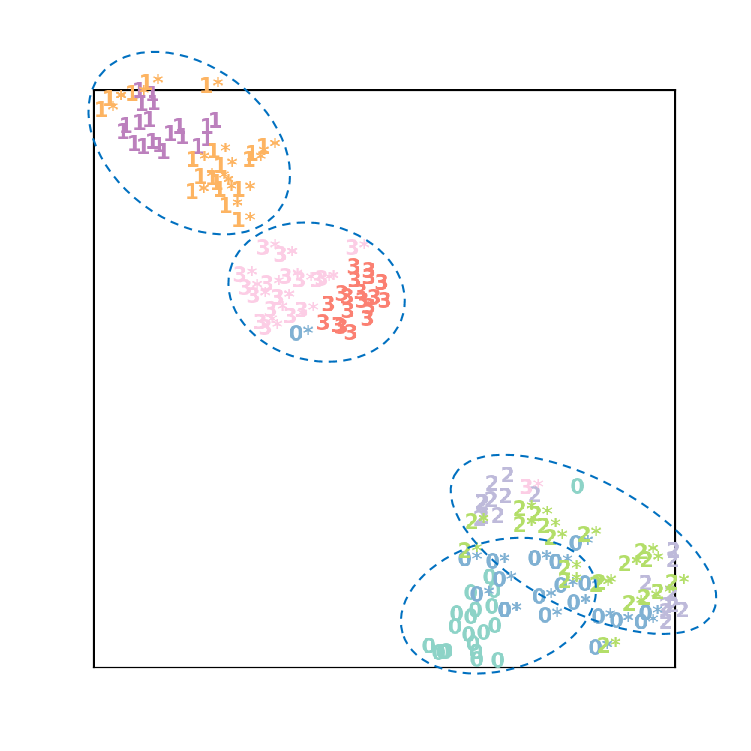}
\end{minipage}%
\begin{minipage}{0.2\textwidth}\label{ssg}
\centering
\includegraphics[width=1\textwidth]{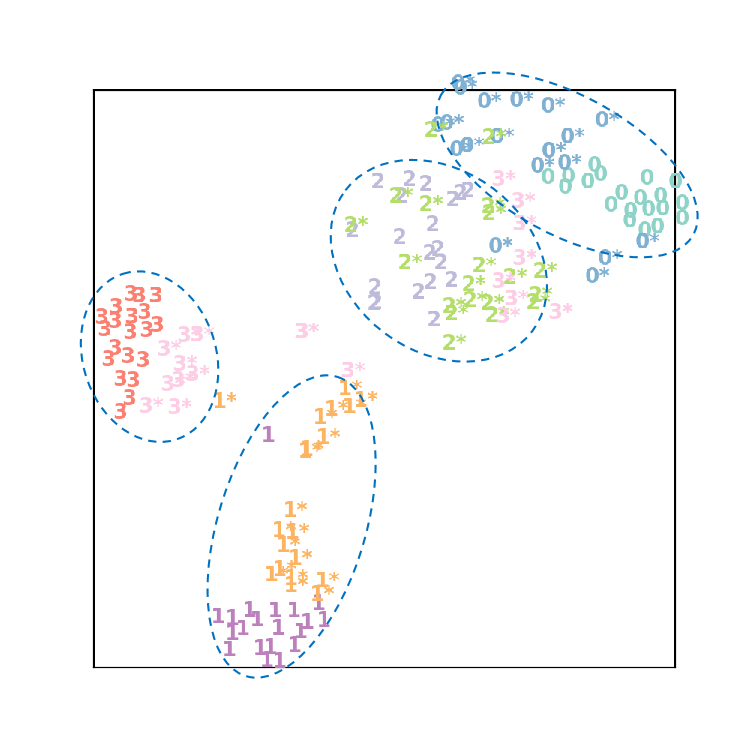}
\end{minipage}%
\begin{minipage}{0.2\textwidth}\label{ssgd}
\centering
\includegraphics[width=1\textwidth]{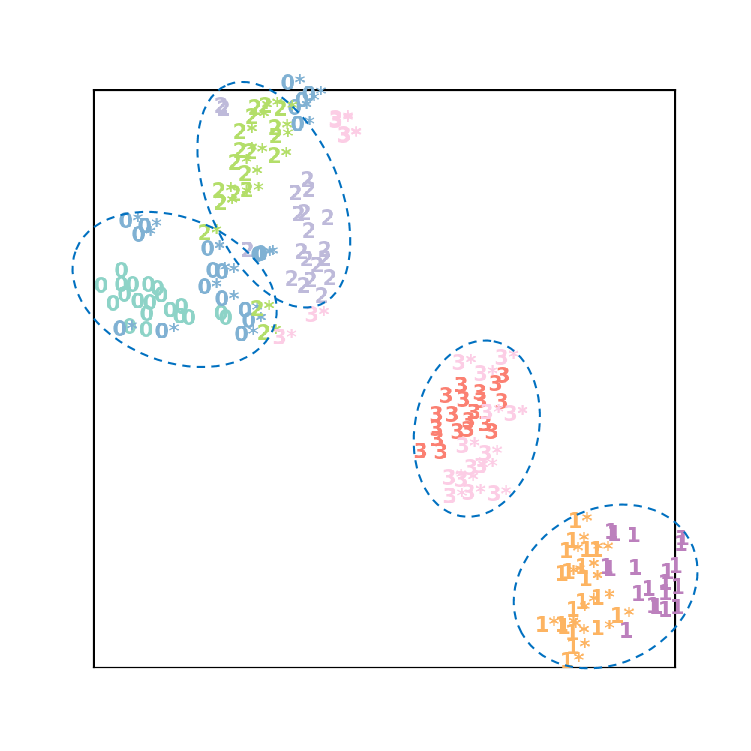}
\end{minipage}%
\begin{minipage}{0.2\textwidth}\label{ssgdc}
\centering
\includegraphics[width=1\textwidth]{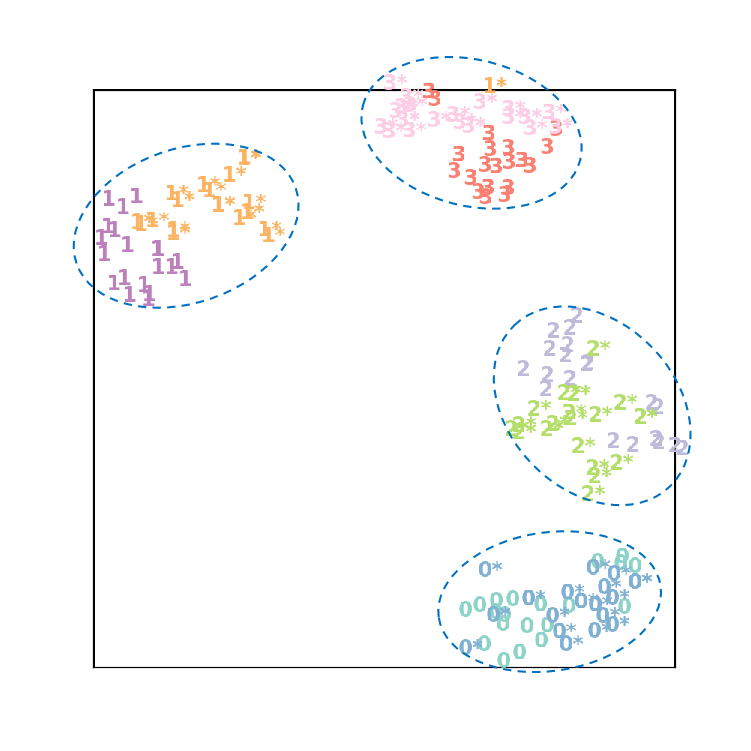}
\end{minipage}
\begin{minipage}{0.2\textwidth}\label{uud}
\centering
\includegraphics[width=1\textwidth]{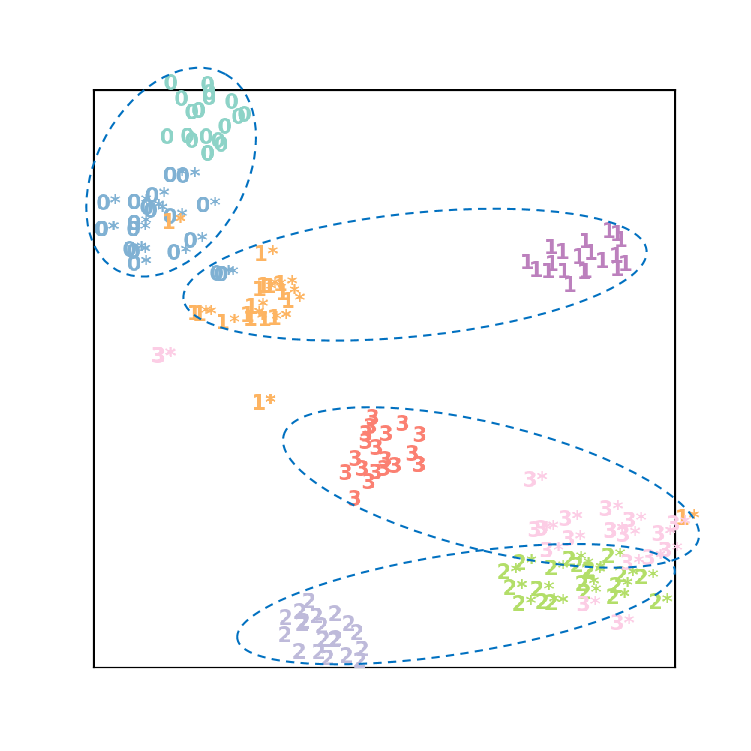}
\end{minipage}%
\begin{minipage}{0.2\textwidth}\label{uug}
\centering
\includegraphics[width=1\textwidth]{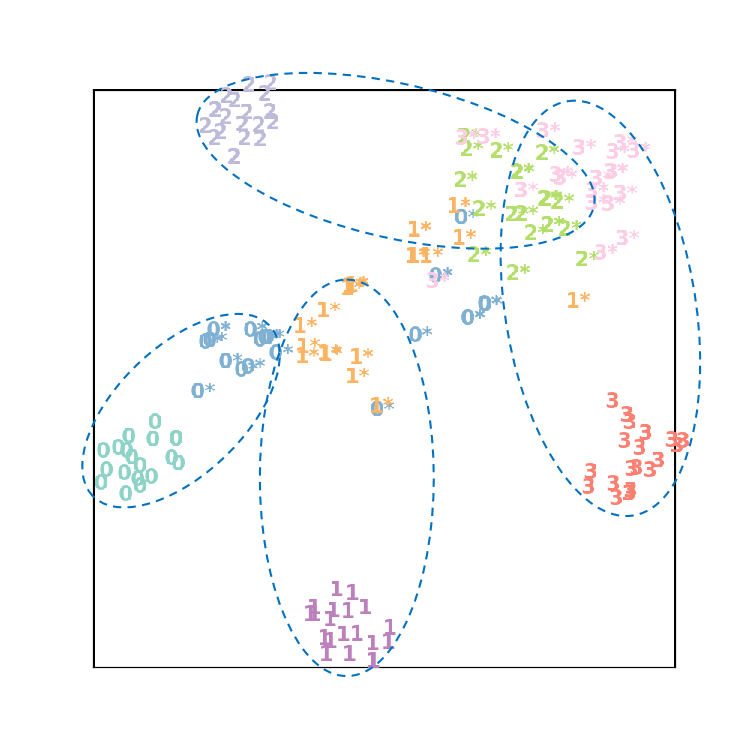}
\end{minipage}%
\begin{minipage}{0.2\textwidth}\label{uugd}
\centering
\includegraphics[width=1\textwidth]{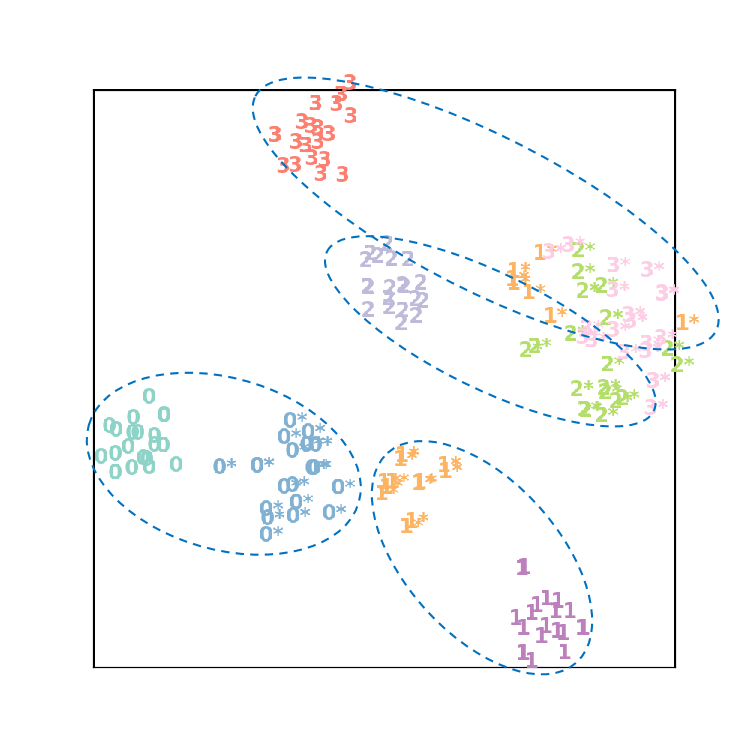}
\end{minipage}%
\begin{minipage}{0.2\textwidth}\label{uugdc}
\centering
\includegraphics[width=1\textwidth]{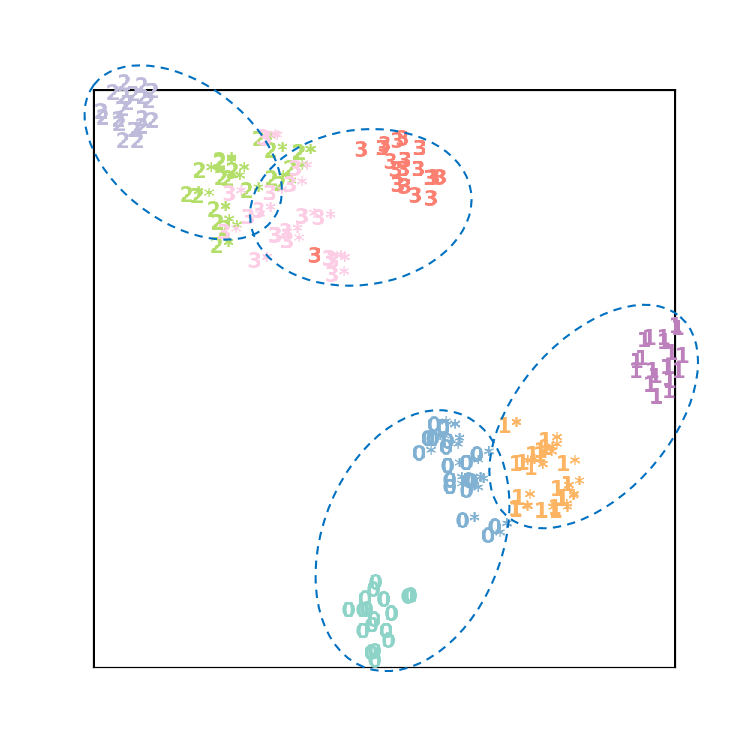}
\end{minipage}
\caption{The visualization of speaker similarity by t-SNE. The upper and lower rows correspond to seen-to-seen VC and zero-shot VC, respectively. From left to right, four columns correspond to D-vector, G-vector, DG-vector and DGC-vector, respectively. Each small number in the graph is an utterance sample. Different numbers without asterisk correspond to different target speakers; numbers with an asterisk refer to converted utterances corresponding to the target speaker whom the number stands for. We roughly mark each conversion pair with a blue dotted circle.}
\label{dv-tsne}
\end{figure*}

\subsection{DGC-vector}
This paper proposes a novel speaker representation for zero-shot voice conversion, which combines the advantages of both the ASV-based method, the GST-based method and auxiliary classification supervision. We call it DGC-vector. Our proposed method consists of three parts: a pre-trained ASV model, a GST-based speaker encoder and an auxiliary classifier.

\subsubsection{D-sequence} \label{D-seq}
D-vector is a widely used kind of ASV method. Although D-vector can achieve good results in speaker recognition, directly using D-vector does not adapt well to zero-shot VC because the pre-trained ASV model and the conversion model are not trained jointly using the conversion optimization function. Inspired by the successful application of phonetic posteriorgrams in VC, we use a pre-trained ASV model to extract the bottleneck features, which we called D-sequence in this paper. The pre-trained ASV model is trained following setting of \cite{wan2018generalized}. As shown in pre-trained ASV model of Fig.~\ref{SE}, the ASV model consists of several LSTM layers and a fully connected (FC) layer. We defined the output of the last LSTM layer as the D-sequence. D-sequence retains time dimension and we believe that D-sequence contains as much speaker information as D-vector and retain more information to adapt to the conversion model. The reason why we choose \cite{wan2018generalized} is that it gets an outstanding performance in ASV task. Indeed, we can flexibly choose other effective method as pre-trained ASV model. The key is that we hope to introduce priori knowledge of ASV task into speaker embedding through pre-trained ASV model.

\subsubsection{GST-based Speaker Encoder}
In order to generate speaker embedding adapted to the VC task, we jointly optimize a GST-based speaker encoder with the conversion model. GST-based speaker encoder has been proved to be effective in extracting a global speaker embeddings for VC~\cite{lu2019one}. This part can also be replaced by other effective models, because what we need is a module which can jointly optimize with conversion model and extract speaker embedding from input feature.
Instead of taking mel-spectrogram as the GST-based speaker encoder's input ~\cite{wang2018style, xie2021multi,lu2019one}, we use the D-sequence described in section~\ref{D-seq} as the input.

The speaker encoder includes two parts, namely a reference encoder $E_{r}$ and a GST module, as shown in Fig.~\ref{SE}.
We assume that the speaker information is time-invariant. Thus we use the reference encoder to extract a fixed-dimension vector. The reference encoder comprises three convolution layers and a GRU\cite{Cho2014LearningPR} layer, and we use the final output of the GRU layer as the input of the GST module. 
A set of randomly initialized learnable vectors is used as GSTs. A multi-head attention module takes the GRU's output as the query and GSTs as the key and value. Through matrix multiplication, we eventually get the final speaker embedding.

\subsubsection{Auxiliary classification task}

To ensure the speaker encoder extracts the speaker-related information, we add an auxiliary classification task as shown in Fig.~\ref{SE}. The classifier includes only one fully connected (FC) layer, which takes the speaker embedding from the GST module as input and predicts its speaker. The classification loss is defined as follows,
\begin{equation}
L_{class}=\mathbb{E}_{x_{ref}\epsilon\chi}\left[-log( C_{a}(c|E_{s}(x_{ref})))\right]
\end{equation}
where $C_{a}$, $E_{s}$ , $\chi$ and $c$ represent auxiliary classifier, entire speaker embedding module, training dataset and speaker identity of input $x_{ref}$, respectively. If we only use the self-reconstruction loss like \cite{lu2019one}, the GST-based speaker encoder may learn some information that is not related to speaker identity~\cite{wang2018style}. Thus it is likely to make the speaker encoder lose its desired function and be harmful for its generalization to the unseen speakers. Therefore, we use the classification task to enforce explicit supervision on the speaker encoder to improve its generalization.

Finally, the objective function of the whole model is defined as
\begin{equation}
L=\lambda_{rec}L_{rec}+\lambda_{class}L_{class}
\end{equation}
where $\lambda_{rec}$ and $\lambda_{class}$ are the weight items for respective loss.

\section{Experiment and evaluation}

\subsection{Dataset and experiment setup}
In the following experiments, we compared different AUTOVC variants with different speaker embeddings, including D-vector, G-vector, DG-vector (DGC-vector without classification supervision), and DGC-vector. 
All models were trained on data of 80 speakers randomly selected from CSTR VCTK corpus~\cite{vckt}. Each speaker has about 400 sentences. We used four target speakers and four source speakers in the seen-to-seen VC scenario. In the zero-shot VC scenario, we used the VCC2020 dataset~\cite{zhao2020voice}. We used TEF1, TEF2, TEM1, and TEM2 as target speakers and SEF1, SEF2, SEM1, and SEM2 as source speakers. 'T'/'S' is for target/source, 'E' is for English, 'F'/'M' is for female/male. In each VC task, we had 16 conversion pairs, including female to female (F2F), male to male (M2M), female to male (F2M), and male to female (M2F). Each model generated 25 converted samples for each pair. 

80-dimension mel-spectrogram is extracted as the input acoustic feature, and the input frame length is 160. Thus the size of input $x$ is 80$\times$160. Speaker embedding generated from different methods is 256-dimension. $\lambda_{rec}$ and $\lambda_{class}$ are set to 1 and 0.5, respectively. Other model configuration settings are the same as that of AUTOVC~\cite{Qian2019ZeroShotVS}. A pre-trained Hifi-GAN~\cite{kong2020hifi} vocoder was used as the vocoder to reconstruct the waveform conditioned on the predicted mel-spectrogram by the conversion model.

\subsection{Objective evaluation} \label{ob}

To visualize speaker similarity of utterances, we plotted the D-vector of the converted utterances and target speakers' utterances in 2D space by t-SNE~\cite{2008Visualizing}. 
As Fig.~\ref{dv-tsne} illustrates, the converted utterances generated by the model with DGC-vector are close to the respective target speakers' utterances in the seen-to-seen case, and different conversion pairs are well separated. In the zero-shot VC, distances between the converted and target utterances are shorter than in other methods. It indicates that the proposed method can improve the speaker similarity and generalization ability in the zero-shot VC task. In addition, the converted utterances gather in a more dense circle, which shows that the speaker consistency of the converted utterances generated by the proposed method is the best. However, the model with G-vector performed well neither in the seen-to-seen scenario nor the zero-shot scenario. As mentioned in section~\ref{d-vec}, the model with D-vector performed significantly worse in the zero-shot scenario. The model with DG-vector suffers from a similar problem.

\begin{table}[tp]
\caption{Result comparison for zero-shot VC among different speaker embedding methods. 'Avg', 'Inter', and 'Intra' mean the average, inter-gender conversion, and intra-gender conversion, respectively.}

\begin{center}
\begin{tabular}{|c|c|c|c|c|c|c|}
\hline
\textbf{Methods}&\multicolumn{3}{|c|}{\textbf{MCD}}&\multicolumn{3}{|c|}{\textbf{MAE of F0}} \\
\cline{2-7} 
  & \textbf{\textit{Avg}}& \textbf{\textit{Inter}}& \textbf{\textit{Intra}}& \textbf{\textit{Avg}}& \textbf{\textit{Inter}}& \textbf{\textit{Intra}} \\
\hline
D-vector& 11.54 & 11.67 & 11.42 & 22.85 & 26.66 & 19.05 \\
G-vector& 11.54 & 11.7 & 11.39 & 22.97 & 28.31 & \textbf{17.63} \\
DG-vector& \textbf{11.32} & \textbf{11.48} & \textbf{11.16} & 22.27 & 26.04 & 18.49 \\
DGC-vector& 11.46 & 11.51 & 11.41 & \textbf{20.5} & \textbf{23.03} & 17.97 \\
\hline
\end{tabular}
\label{tab1}
\end{center}
\end{table}

Dynamic time warping Mel cepstral distortion (DTW-MCD)~\cite{407206} and mean absolute error (MAE) of F0 were used as the objective metric, and the lower the better. We used the converted utterances and the ground truth data to calculate the metrics. Because we only got the ground truth data for the VCC2020 dataset, we only computed the results for zero-shot VC. The results are shown in Table~\ref{tab1}. In terms of MCD,  the proposed method gets a slight improvement over D-vector and G-vector. For MAE of F0, the proposed method is significantly lower than other methods, which indicates that the proposed method has a better F0 similarity which is related to voice characteristics of speakers.

\subsection{Subjective evaluation}

\iffalse
\begin{figure}[tp]
\centering
\includegraphics[width=0.5\textwidth]{fig/uu-res_nature.pdf} 
\caption{MOS of naturalness for zero-shot VC. Error bars denote the standard deviation of samples.}
\label{na}
\end{figure}

\begin{figure}[tp]
\centering
\includegraphics[width=0.5\textwidth]{fig/uu-res_similar.pdf} 
\caption{MOS of similarity for zero-shot VC. Error bars denote the standard deviation of samples.}
\label{si}
\end{figure}
\fi

%\iffalse
\begin{figure}[tp]
\centering
\begin{minipage}{0.5\textwidth}
\centering
\includegraphics[width=1\textwidth]{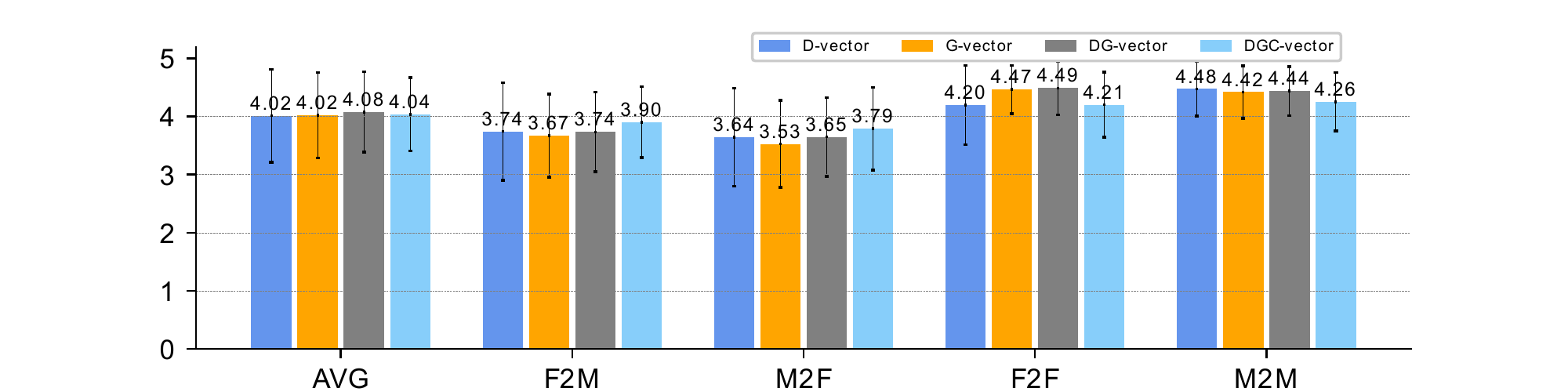} 
\end{minipage}
\begin{minipage}{0.5\textwidth}
\centering
\includegraphics[width=1\textwidth]{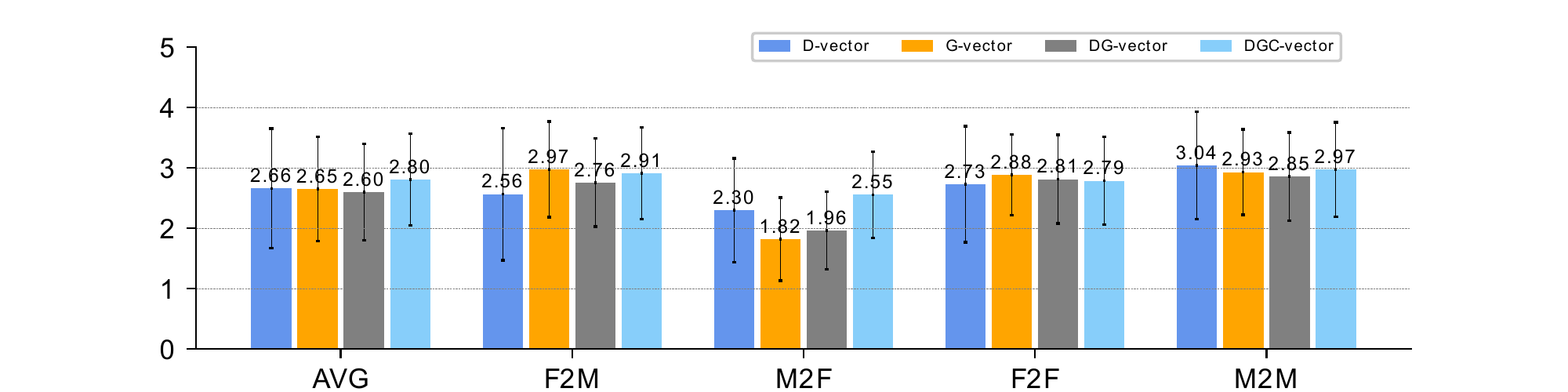} 
\end{minipage}
\caption{MOS of naturalness (upper) and similarity (bottom). Error bars denote the standard deviation of samples.}
\label{si}
\end{figure}
%\fi

%For the subjective evaluation, we used the mean opinion score (MOS) test. We used 5 randomly selected utterances of 25 generated converted samples for each conversion pair, so each method had 160 (2$\times$16$\times$5) utterances in all. Thirteen Chinese raters who are proficient in English evaluated the naturalness and speaker similarity of utterances using scores ranging from 1 to 5; the higher the better. The samples can be found at \href{https://shaw0fr.github.io/DGC-vector-DEMO/}{DEMO}.
For the subjective evaluation, we used the mean opinion score (MOS) test in zero-shot VC scenario. We used 5 randomly selected utterances of 25 generated converted samples for each conversion pair, so each method had 80 (16$\times$5) utterances in all. Thirteen Chinese raters who are proficient in English evaluated the naturalness and speaker similarity of utterances using scores ranging from 1 to 5; the higher the better. The demo can be found at \href{https://shaw0fr.github.io/DGC-vector-DEMO/}{https://shaw0fr.github.io/DGC-vector-DEMO/}.

%In the naturalness MOS test, different methods gave comparable performances in both seen-to-seen VC and zero-shot VC as shown in Fig.~\ref{na}. This result is unsurprising since we only make modifications to the speaker representation. What is surprising is that the performance of zero-shot VC is quite comparable to seen-to-seen VC. We suspect that our raters are not native speakers and not aware of nuance in the converted speech.
%In the naturalness MOS test, we only evaluated methods for zero-shot VC, because we need to reduce the workload of raters and paid more attention to zero shot and similarity. We noticed that different methods gave comparable performances as shown in Fig.~\ref{na}. This result is unsurprising since we only make modifications to the speaker representation. What is surprising is that the score are much higher than previous work~\cite{Qian2019ZeroShotVS}. We suspect that our raters are not native speakers and not aware of nuance in the converted speech.
In the naturalness MOS test, we notice that different methods give comparable average performances as shown in the upper part of Fig.~\ref{si}. This result is unsurprising since we only made modifications to the speaker representation. However, we find that the naturalness MOS of D-vector, G-vector and DG-vector get worse during inter-gender VC. The problem may be related to generalization ability of methods. And DGC-vector is more stable. What is surprising is that the score are much higher than previous work~\cite{Qian2019ZeroShotVS}. We suspect that our raters are not native speakers and not aware of nuance in the converted speech.

%In the speaker similarity MOS test, we evaluated speaker similarity in the seen-to-seen VC task and the zero-shot VC task. And the result is shown in Fig.~\ref{si}. For the seen-to-seen VC task, our method got the highest score. Compared with G-vector and DG-vector in seen-to-seen VC, we find that D-sequence is superior to mel-spectrogram as the input of the GST-based speaker encoder. We suspect that the GST-based speaker encoder may not remove all the extra information in the mel-spectrogram. However, G-vector and DG-vector provided similar performance in zero-shot VC, revealing that GST-based speaker encoder alone fails to process the "unseen" D-sequence. Compared with D-vector and DG-vector, DG-vector did not give much improvement in both tasks, which indicates the unsupervised GST speaker encoder does not leverage the extra helpful information in D-sequence. With the classification supervision, speaker similarity for both tasks significantly increased, which indicates DGC-vector has the best generalization ability. Moreover, DGC-vector is more stable than other methods. Our experiments show that the improvement of DGC-vector effect is not simply the accumulation of positive benefits of multiple modules, because the improvement of DG-vector is not obvious. In order to make full use of D-sequence and GST, auxiliary supervision is needed. Only when the three parts work together can we improve the speaker similarity and generalization ability.
In the speaker similarity MOS test, we evaluated speaker similarity in the zero-shot VC task. As the bottom part of Fig.~\ref{si} declares, our method got the highest average score. Compared with G-vector and DG-vector, we find that D-sequence does not bring expected benefit, although the performance of G-vector is more unstable for different conversion pairs and significantly gets worse for male-to-female VC. We suspect that the GST-based speaker encoder may not make full use of the speaker information contained by D-sequence. Compared with D-vector and DG-vector, DG-vector did not give much improvement, which also indicates the unsupervised GST speaker encoder does not leverage the extra helpful information in D-sequence. However, with the classification supervision, speaker similarity remained relatively high score for different conversion pairs, which indicates DGC-vector has the best generalization ability. Moreover, DGC-vector is more stable than other methods, similar to the situation in naturalness MOS test. Especially during inter-gender VC, whether it is male-to-female or female-to-male, our method can give a good performance.  These conclusions are consistent with the conclusions obtained from Fig.~\ref{dv-tsne} in section~\ref{ob}. Our experiments show that the improvement of DGC-vector effect is not simply the accumulation of positive benefits of multiple modules, because the improvement of DG-vector is not obvious. In order to make full use of the advantages of D-sequence and GST, auxiliary supervision is needed. 
%Only when the three parts work together can we improve the speaker similarity and generalization ability.
Conclusively, although results of naturalness MOS are similar, our method can significantly improve the speaker similarity and generalization ability.

\section{Conclusions}
In this paper, we study the impact of speaker embedding on zero-shot VC effects. In order to represent the characteristics of target speaker better and improve speaker similarity of zero-shot VC, we propose a novel speaker embedding method. Under the framework of AUTOVC, a GST-based module with auxiliary supervision is applied to generate speaker embedding from bottleneck feature of ASV model. Objective and subjective evaluations show that the proposed method can improve speaker similarity and generalization ability.

\bibliographystyle{IEEEbib}
\bibliography{dgc}

\end{document}